\documentclass[letterpaper,twocolumn,10pt]{article}
\usepackage{usenix2019_v3}

\usepackage{enumitem}
\usepackage{amsmath}
\usepackage{amssymb, amsthm}
\usepackage{stmaryrd}

\usepackage{caption}
\usepackage{graphicx}
\usepackage{subcaption}

\theoremstyle{definition}

\newcommand{\ssdeep}{\texttt{ssdeep}}

\usepackage{algorithmic}
\usepackage[noend,lined,linesnumbered]{algorithm2e}

\usepackage{xcolor}
\definecolor{teal}{HTML}{008080}

\usepackage{hyperref}
\hypersetup{
   colorlinks=true,
   citecolor=teal
}

\usepackage{multirow}

\allowdisplaybreaks[1] 

\newcommand{\descr}[1]{\vspace{0.2cm} \noindent \textbf{#1}}

\title{On the Robustness of Malware Detectors to Adversarial Samples\thanks{This is the full version of the paper with the same title to appear in the proceesdings of the 2024 Workshop on Security and Artificial Intelligence (SECAI 2024).}}

\author{\rm Muhammad Salman,
Benjamin Zi Hao Zhao, Hassan Jameel Asghar\\ 
\rm Muhammad Ikram, Sidharth Kaushik \&
Mohamed Ali Kaafar \\
\small \rm Macquarie University, Australia\\
\small \rm \texttt{\{muhammad.salman, ben\_zi.zhao, hassan.asghar,}\\
\small \rm \texttt{muhammad.ikram, sidharth.kaushik, dali.kaafar\}@mq.edu.au
}
}

\date{}
\begin{document}

\maketitle

\begin{abstract}
Adversarial examples add imperceptible alterations to inputs with the objective to induce misclassification in machine learning models. They have been demonstrated to pose significant challenges in domains like image classification, with results showing that an adversarially perturbed image to evade detection against one classifier is most likely transferable to other classifiers. Adversarial examples have also been studied in malware analysis. Unlike images, program binaries cannot be arbitrarily perturbed without rendering them non-functional. Due to the difficulty of crafting adversarial program binaries, there is no consensus on the transferability of adversarially perturbed programs to different detectors. In this work, we explore the robustness of malware detectors against adversarially perturbed malware. We investigate the transferability of adversarial attacks developed against one detector, against other machine learning-based malware detectors, and code similarity techniques, specifically, locality sensitive hashing-based detectors. Our analysis reveals that adversarial program binaries crafted for one detector are generally less effective against others. We also evaluate an ensemble of detectors and show that they can potentially mitigate the impact of adversarial program binaries. Finally, we demonstrate that substantial program changes made to evade detection may result in the transformation technique being identified, implying that the adversary must make minimal changes to the program binary.
\end{abstract}

\section{Introduction}
Adversarial examples are inputs with slight, imperceptible modifications that induce misclassification in machine learning models. The efficacy of this attack has been well documented in the domain of image classification where perturbed images that are otherwise human-recognizable are misclassified by victim machine learning models \cite{carlini2017towards, zhang2020adversarial}. A series of recent works has also highlighted this vulnerability in machine learning-based malware detection \cite{dahl2013large, grosse2016adversarial, tramer2017space, suciu2019exploring, hu2022transferability}. A key difference between the two domains is that while slight changes to an image, e.g., by injecting white noise, still deems it recognizable (by a human), arbitrary changes to a program can render it dysfunctional. One major consequence of this difference in the study of adversarial examples is that while in the domain of images, adversarial example attacks are largely transferable~\cite{tramer2017space}, meaning adversarial examples crafted for a target machine learning model can be used to fool a different machine learning model, this has been shown to be largely false in the malware detection domain~\cite{suciu2019exploring}.\footnote{Some success has been reported~\cite{hu2022transferability} but with a feature space that is highly restricted, e.g., a binary feature space where we can append dead code without affecting the functionality of the program~\cite{grosse2017adversarial}.} 

This lack of transferability has profound implications for adversarial example attacks in malware. For instance, real-world systems rarely rely on a single malware detector. Consider for instance, the VirusTotal service.\footnote{See \url{https://www.virustotal.com/}} Given a program binary or its cryptographic hash, it returns a detailed report on the malicious nature of the program. Included in this report are its cryptographic hash (e.g., SHA-256), locality sensitive hash (LSH) digests (e.g., \ssdeep{}~\cite{ssdeep}), as well as detection results from various security vendors (scanners) some of whom employ machine learning-based detectors~\cite{lucas2021malware}. Thus, a malware program adversarially perturbed to evade a particular target machine learning model will not evade detection against an ensemble of detectors. To understand why an adversarial example attack on a machine learning model may not be transferable to a machine learning-based or non-machine learning-based detector consider an LSH algorithm for malware detection. Roughly, such an algorithm divides its input into blocks, computes the hash of each block and concatenates the results to create the final digest. Thus, if a certain block is unchanged, the hash of that block will remain the same. One of the techniques to create adversarial examples is to append code to the end of the program~\cite{ebrahimi2021binary}. Since this is only added to the end of the program, the LSH-based detection scheme will show the same digest for the blocks of the program before the appended part, and hence the program will still be detected as malware.

A naive way to solve this issue, from the attacker's perspective, would be to completely obfuscate the program, using packers for example. However, the presence of obfuscation may itself result in the program being flagged as potential malware~\cite[\S 5]{pierazzi2020problemspace}. Indeed, packing with popular packers increases likelihood of being detected~\cite[\S 4.4]{lucas2021malware}. Also, obfuscation may also induce large changes. Too many changes are not ideal as they would create new signatures that could be used to identify the technique used to perturb a program~\cite[\S 4]{anderson2018learning}. Therefore, in the spirit of adversarial examples in the image domain, it may make sense to only consider the attack potent if the amount of changes made to the code are minimal. 

In this paper, we empirically investigate these assertions. More concretely, our contributions are as follows:

\begin{itemize}
    \item We empirically demonstrate that an adversarial example attack targeted on one detector is not in general transferable to another. To demonstrate this, we take a recently proposed framework called malware makeover~\cite{lucas2021malware}, and use it to transform program binaries to evade detection against a deep neural network (DNN) based model trained on raw bytes (MalConv~\cite{raff2018malconv}), as is done in~\cite{lucas2021malware}. We then test these transformed binaries against the original model, i.e., MalConv, an \ssdeep{} detector (a commonly used LSH algorithm), and a random forest (RF) based detector trained on the EMBER dataset~\cite{ember_dataset} with features via LIEF~\cite{LIEF},\footnote{The features obtained from binaries are known as EMBER features} a set of features automatically extracted from program binaries. Our results show that the Malconv targeted transformed binaries are poor in evading detection against the other detectors. This is rather surprising, since as we show in Section~\ref{sub:ssdeep}, evading \ssdeep{} detection is not difficult. We repeat the same procedure with taking these other detectors as the target for transforming programs and show that the results stay the same.   
    \item We then investigate whether an ensemble of detectors is capable of mitigating the impact of adversarially transformed binaries, as the above conclusion may imply, using different decision rules: minority, majority and consensus. We take all three detectors as part of the ensemble and report the impact on the true/false positive rates of the original binaries, and the impact on these rates if the binaries are transformed taking one of these detectors as the target detector to create adversarial samples. Our results indicate that a majority voting strategy is capable of substantially mitigating this adversarial attack since its success rate is substantially lower on the non-target detectors. We show that this conclusion is also backed by results from the VirusTotal service.
    \item We show that if programs are substantially changed, then it creates a new signature of the underlying scheme. This has been raised as a possible issue with transformation techniques that change the program too much~\cite[\S 4]{anderson2018learning}. We validate this by training a model on original versus modified binaries to see if it is able to detect the modified program via the malware makeover technique~\cite{lucas2021malware}. An implication of this is that an adversary who seeks to evade detection against an ensemble of detectors would need to make substantially more changes to the original program resulting in the program being detected as heavily obfuscated.
\end{itemize}

\section{Preliminaries}

In this section we provide necessary background with definitions, our threat model, workings of the malware makeover technique, locality sensitive hashing in the context of an adversarial attack, and the engineered malware analysis features of EMBER.

\subsection{Definitions}
\label{sub:defs}
We assume that we have a set $D = B \cup M$ of benign $B$ and malware $M$ programs and their labels. Each program $x \in D$ has the label $0$ if $x \in B$, else it has the label $1$. The two subsets $B$ and $M$ of $D$ are disjoint.

\descr{True and False Positives.} Given a malware detector $A$, we define its type-I (false positive) and type-II (false negative) error rates as:
\[
\text{FPR}_A = \frac{|A(x) = 1 : x \in B|}{|B|}, \; \text{FNR}_A = \frac{|A(x) = 0 : x \in M|}{|M|}
\]
Upon applying an evasive technique $\mathcal{E}$ on the programs in $D$, we define the same types of errors as:
\begin{gather*} 
\text{FPR}_A^{\mathcal{E}} = \frac{|A(x') = 1 : x' \leftarrow \mathcal{E}(x), x \in B|}{|B|}, \\ 
\text{FNR}_A^{\mathcal{E}} = \frac{|A(x') = 0 : x' \leftarrow \mathcal{E}(x), x \in M|}{|M|}
\end{gather*}
The changes in FPR and FNR measure the evasiveness of the technique. Instead of FNR, we shall report the true positive rate (TPR), which is the oppositive of it, i.e., $\text{TPR} = 1 - \text{FNR}$.

\descr{Program Similarity.} We take the definition of binary code similarity from~\cite{marcelli-ml-binary}. That is, we say that two binaries are similar if they are compiled from two source codes that are similar. We are also interested in the similarity of the original program $x$ and its evasive counterpart $x' \leftarrow \mathcal{E}(x)$. We assume there to be a distance metric $\mu_0$ that captures this. We shall use the normalized Levenstein distance (NLD) as this metric~\cite{yujian2007normalized}.

\descr{Machine Learning vs Hash-based Detection.} For machine learning-based detection, we assume a binary classifier $f$, which takes a program binary $x$ as input and outputs 1 or 0, indicating malware or benign, respectively. On the other hand, for hash-based detection we assume a signature-based scheme. One such family of hash functions is locality-sensitive hashing (LSH). The hash-based detection scheme is endowed with the triplet $(H, \mu, \theta)$, respectively, a hash function, a metric on the hashes, and a threshold. The metric $\mu$ on the hashes takes as input two hashes $h' = H(x')$ and $h'' = H(x'')$ and outputs $\mu(h', h'')$. Given a hash $h$ and a set of signatures $S$, we define $\text{dist}_\mu(h, S) = \min_{h' \in S} \mu(h, h')$. The hash based scheme works as follows:

\begin{enumerate}
    \item In the training phase, for each $x \in D$, i.e., malware or benign program in the dataset, it calculates $H(x)$ and stores it in the database $S$. These are called \emph{signatures}. 
    \item Given a program $x$, it calculates its hash $H(x)$. 
    \item It computes $d = \text{dist}_\mu(h, S)$. If $d \leq \theta$, it outputs 1 (malware), else it outputs 0 (benign).
\end{enumerate}
In the one extreme, $h$ could be a cryptographic hash function, in which case we can set $\theta = 0$, and the program outputs $1$ if the hash is identical to some signature in the database. Note that the distance function in the hash domain may not be the same metric used to calculate similarity of the original programs. Depending on the situation, there may be multiple programs that pass the test of Step 3. In this case, we only consider the detector as successful if the only program to pass the test of Step 3 when given $h = H(\mathcal{E}(x))$ is $x$, i.e., the original program. 

\descr{Threat Model.}  The adversary would like to make changes to a program so that it is misclassified by the target detector while retaining functionality. As discussed in the introduction a criterion for the transformed program is that it should remain similar to the original program in terms of edit distance, e.g., according to the NLD metric $\mu_0$ above. Given this, the goal of the adversary is as follows: Given the set $D = B \cup M$ of benign and malware programs, a detector $A$ with false positive and true positive rates of $\text{FPR}_A$ and $\text{TPR}_A$, respectively, a distance metric $\mu_0$ and a distance threshold $\theta_0$, construct an evasive technique $\mathcal{E}$ such that:
\begin{enumerate}
    \item For all $x \in D$, $\mu_0(x, x') \le \theta_0$, where $x' \leftarrow \mathcal{E}(x)$
    \item $\text{FPR}^{\mathcal{E}}_A > \text{FPR}_A$  and $\text{TPR}^{\mathcal{E}}_A < \text{TPR}_A$.
\end{enumerate}



\subsection{Malware Makeover and Extension}
\label{sec:mmo}
We first describe the algorithm behind the malware makeover attack from~\cite{lucas2021malware}. 

\descr{White-Box Attack.} The white-box malware makeover attack targets a machine learning detector which uses gradient descent for minimizing its loss, e.g., a neural network. Let $\mathbf{b}$ denote a raw binary. We assume an embedding function $E$ that takes a binary $\mathbf{b}$ and outputs a feature vector $\mathbf{x}$. The white-box malware makeover algorithm takes a function $f \in \mathbf{b}$, and transforms it to produce a new binary $\mathbf{b}'$. Let $E(\mathbf{b}') = \mathbf{x}'$. These transformations will be introduced shortly. The  transformed binary can be represented as $ \mathbf{x}' = \mathbf{x} + \boldsymbol{\delta}$. If the resulting vector increases the loss of the detector then the transformation is retained, otherwise it is rejected. We can view the \emph{displacement vector} as $\boldsymbol{\delta} = \alpha g(\mathbf{x})$. Here $\alpha > 0$ is a small scalar, and $g$ is the gradient w.r.t. input $\mathbf{x}$. Let $\ell(\mathbf{x})$ denote the loss on input $\mathbf{x}$. Then using Taylor's approximation:
\begin{align*}
    \ell({\mathbf{x} + \boldsymbol{\delta}}) &\approx \ell(\mathbf{x}) + \langle g(\mathbf{x}), \boldsymbol{\delta}\rangle \\
    &= \ell(\mathbf{x}) + \langle g(\mathbf{x}), \alpha g(\mathbf{x}) \rangle \\
    &= \ell(\mathbf{x}) + \alpha \langle g(\mathbf{x}), g(\mathbf{x}) \rangle\\
    &= \ell(\mathbf{x}) + \alpha  \lVert g(\mathbf{x}) \rVert_2^2 \\
    &> \ell(\mathbf{x}). 
\end{align*}
Thus, after obtaining $\mathbf{x}'$, we check if the dot product $\langle g(\mathbf{x}), \boldsymbol{\delta}\rangle > 0$, where $ \boldsymbol{\delta} = \mathbf{x}' - \mathbf{x}$. If this is true, then the loss has increased. In other words, we are travelling in the direction of the gradient to increase loss. At some point the loss becomes high enough for the detector to misclassify, at which point the algorithm stops. 

\descr{Black-Box Attack.} Let us assume that we have the cross-entropy loss. Then if $K$ denotes the number of classes and $c$ is the index of the correct label (one-hot encoded), we have:
\[
\ell(\mathbf{x}) = - \sum_{i = 1}^K y_i \ln p_i(\mathbf{x}) = - \ln p_c(\mathbf{x}).
\]
Rearranging we get 
\[
e^{- \ell(\mathbf{x})} = p_c(\mathbf{x}).
\] 
Thus, we can instead check if $p_c(\mathbf{x}') < p_c(\mathbf{x})$. If that is the case then necessarily $\ell(\mathbf{x}') = \ell({\mathbf{x} + \boldsymbol{\delta}}) >  \ell(\mathbf{x})$. The advantage is that this makes it a blackbox attack rather than a whitebox attack, since we do not need to compute the gradient. The above should work for any loss function since:
\[
\ell(\mathbf{x}) \propto \frac{1}{p_c(\mathbf{x})},
\]
Hence this attack can be applied to models other than neural networks (which usually employ gradient descent). This black-box attack is also mentioned in~\cite{lucas2021malware}.

The main point where the white-box attack from~\cite{lucas2021malware} deviates from its black-box counterpart is where we can set arbitrary integer values to \emph{semantic nops}. To use an example from~\cite{lucas2021malware}, suppose the integer value of the semantic nop corresponds to the $i$th byte in the binary $\mathbf{b}$. We replace the $i$th byte of $\mathbf{b}$ by $b \in \{0, 1, \ldots, 255\}$, such that the dot product $\langle g(E(\mathbf{b})), \boldsymbol{\delta}\rangle$ is maximized, where $\boldsymbol{\delta} = E(\mathbf{b}') - E(\mathbf{b})$ and
\begin{align*}
    \mathbf{b} = (b_1, b_2, \ldots, b_i, \ldots, b_n) \\
    \mathbf{b}' = (b_1, b_2, \ldots, b', \ldots, b_n)
\end{align*}
for a binary of length $n$ bytes. However, we could do the same attack in the black-box setting by replacing the $i$th byte of $\mathbf{b}$ by every possible byte value $b \in \{0, 1, \ldots, 255\}$ and choosing the resulting binary $\mathbf{b}'$ that results in the least probability $p_c (E(\mathbf{b}'))$. Of course the drawback here is that we need to do inference a total of 256 times for each such semantic nop (instead of obtaining this information through the gradient in the white-box version of the attack).

\subsubsection{Transformations}
\label{subsub:transformations}
Malware makeover makes use of two families of transformations, which were initially proposed to enhance security of program binaries and deter code-reuse attacks. The first family introduces in-place randomization (IPR) \cite{pappas2012smashing} strategies that preserve the functionality of binaries while aiming to obscure their structure from potential attackers. This methodology entails four distinct transformation types: 1) substituting instructions with functionally equivalent counterparts without altering their length; 2) reassigning registers within specific scopes, ensuring subsequent code remains unaffected; 3) reordering instructions based on dependencies to maintain execution order; and 4) modifying stack operations to preserve register values across function calls, all while conservatively avoiding speculative disassembly to ensure code semantics are retained.
However, the initial IPR approach had limitations in generating the range of functionally equivalent binary variants theoretically possible under these transformation types. To address these limitations, Lucas et al ~\cite{lucas2021malware} enhanced the implementation to allow iterative application of transformations to the same function, apply transformations more conservatively (especially around instructions affecting the flags register), and support a broader set of instructions and calling conventions. This enhancement aimed to better maintain the binary's functionality while broadening its transformation capabilities. A simple example of an IPR transformation (substituting equivalent instruction) is~\cite{lucas2021malware}: 
 \texttt{add ebx 0x10 --> sub ebx, -0x10}.   

The second family of transformations, derived from the concept of Displacement (Disp) \cite{ koo2016juggling}, focuses on disrupting potential attack vectors by displacing code to a new executable section. This strategy involves replacing segments of the original code with a jump instruction to the new location and updating any relative addresses to maintain functionality. To improve upon this, the authors extended Disp by enabling the displacement of any set of consecutive instructions and substituting original instructions with semantic no-operations (nops) -  instruction sequences that collectively have no impact on memory or register states and produce no side effects \cite{christodorescu2005semantics}. To prevent the size of the resultant binary from growing above a predetermined threshold (such as 1\% above its original size), the attack begins by displacing code up to a specific budget (amount of bytes to be displaced). The number of functions in the binary is divided by the budget, and we try to move precisely that many bytes per function. We make a random selection if there are several possibilities for replacing a function's code. Our assessments expand upon the transformation framework and codebase introduced in Malware Makeover~\cite{malmakeover}, including its {\it private} version--acquired direct correspondence with the authors of \cite{lucas2021malware}. Our extensions to this codebase are detailed in Section~\ref{sub:transformed-binaries}. 

\subsubsection{Excluding Headers}
\label{subsub:exclude-header}
We exclude the header during the training and classification of binaries in our malware detection model. This decision is motivated predominantly from the malware makeover work~\cite{lucas2021malware} who also strip the headers. This is due to the finding in~\cite{ demetrio2019explaining} that deep neural networks (DNNs) when applied to binary files, tend to base their classification on characteristics found within the file header rather than meaningful features from the data and text sections where malicious content typically resides. Adversaries can craft malware binaries with minimal header modifications to evade detection, underscoring the need for a model that focuses on more substantive features from the body of the program.

\subsection{Locality Sensitive Hashing and Adversarial Example Attack}
\label{sub:ssdeep}

Locality sensitive hash (LSH) functions roughly apply a hash function on disjoint chunks of the input. The idea is that unless enough chunks have been changed, the resulting digest will look similar to the original digest. This is unlike cryptographic hash functions. For a representative of LSH, we take \ssdeep{}, which is a popular algorithm used in malware detection. We provide a simplified description of \ssdeep{}. Given a string $s$, the \ssdeep{} algorithm uses a rolling window of a specified length (in terms of bytes). At each byte sequence covered by the rolling window, it computes a rolling hash. If the rolling hash of the current byte sequence (determined by the window) is equal to $-1 \pmod{b}$, then this triggers the algorithm to compute the cryptographic hash of the current chunk, i.e., all byte sequences until now. The 6 least significant bits of this hash are retained. These bits are then converted into a Base64 representation, resulting in the Base64 hash character of the \ssdeep{} hash of the current chunk. The window then moves to the next byte sequence until the next ``trigger point'' is reached, and the next Base64 hash character is appended to the previous hash character. A total of 64 chunks are used in the default setting. The modulus $b$ is refered to as the block size and satisfies the approximate relation: $b \approx \frac{n}{S}$, where $n$ is the file size in bytes, and $S = 64$ denotes the number of chunks~\cite{kornblum-ssdeep, baier-ssdeep-attack}. If the number of chunks are fewer than $S/2 - 1 = 31$, the algorithm sets the blocksize to $b/2$, and the process is repeated. 

Below is an example of the \ssdeep{} hash of a string of length 500:
\begin{verbatim}
12:+xK3x+44q49oMeCsavrOyp9YQRHqHsaQCvuvhn+Q
yVyjfA5:j3X4q4DeCsaviypeQRH4LQFcQyj    
\end{verbatim}

\begin{figure*}[th]
\centering
\includegraphics[width=1\linewidth]{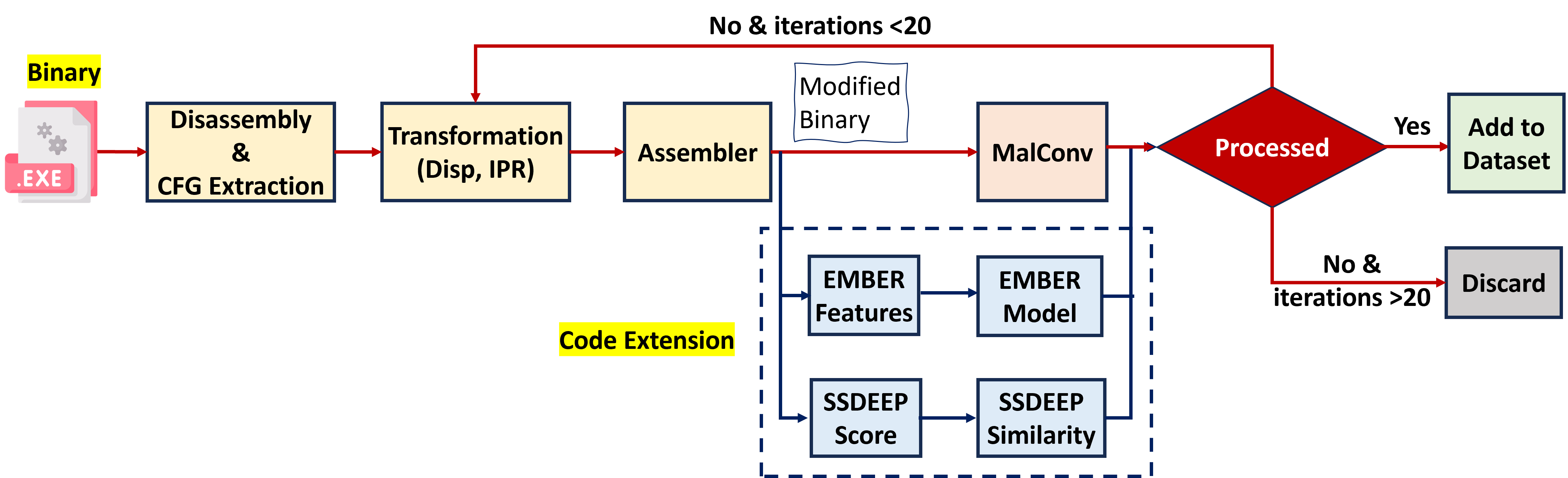}
\caption{Pipeline for functionality preserving transformation of binaries using Malware Makeover. }
\label{fig:process}
\end{figure*}

The last substring starting from the `\verb+:+' is the \ssdeep{} hash with the block size $12 \times 2 = 24$. Since the number of Base64 characters is less than $31$, the algorithm sets the block size to $12$ (the integer at the start of the hash), and computes the hash shown between the two \verb+:+'s. 

\descr{Adversarial Example Attack on \ssdeep{}.} One of the attacks on \ssdeep{} reported in~\cite{baier-ssdeep-attack} when the input is natural language text is to change one byte in each chunk. To ensure that trigger points are not impacted, the last 7 bytes in each chunk are left untouched (corresponding to the window size). In fact, by utilising the fact that the \ssdeep{} similarity checking algorithm only considers a match if there is one common substring in the two hashes of length 7~\cite{baier-ssdeep-attack}, even less number of changes need to be made. A simpler version of this attack, that makes changes to a character in every substring of size $b \approx n/64$, seems to work nearly 100 percent of the time. Thus, we need not even identify chunks. This can be explained as follows. The rolling hash identifies the end of a chunk if the current byte sequence evaluates to $-1 \pmod{b}$. If the rolling hash is pseudorandomly distributed, then the probability that the current sequence of bytes evaluates to this is $1/b$. The probability that a chunk has not been identified in the first $7b$ rolling windows is given by:
\[
\left(1 - \frac{1}{b}\right)^{7b} \approx e^{-7} = 0.00090359397765.
\]
This is very low, and hence the probability that the simpler attack will work is high. 

Implementing this attack in the malware domain requires changes in each chunk of a given binary file. To maintain functionality, these changes should be meaningful. One such example could be using code substitution. For instance, among the different transformation types mentioned in the work from Lucas et al.~\cite{lucas2021malware} to evade a target machine learning malware detector is replacing instructions with equivalent ones, which preserve length. For example, \verb|sub eax,4| can be replaced by \verb|add eax,-4|. This is an example of an IPR transformation. If each block contains such an instruction, we can replace them with equivalent ones and be able to evade detection via \ssdeep{}.

\subsection{EMBER Features}
For one of our detectors, we use the EMBER feature set, which is a set of engineered features extracted from software binaries commonly found in benchmark evaluation datasets first introduced by EMBER~\cite{ember_dataset} and reinforced by SOREL~\cite{harang2020sorel20m}, containing 900K and 20M software binaries respectively.
The EMBER features are structured into nine distinct feature categories, encapsulating both parsed heuristics and format-agnostic statistical measures. To ensure that the features are not extracted from the header of the binary file (see Section~\ref{subsub:exclude-header}), we only extract 
features from the data and text sections. Thus, we only retain ByteHistogram, ByteEntropy, StringExtractor, and SectionInfo features, excluding GeneralFileInfo, HeaderFileInfo, ImportsInfo, ExportsInfo, and DataDirectories. Appendix~\ref{app:ember_feat} provides a brief description of each extracted feature.

\section{Experimental Setup}
In this section we describe the evaluation datasets, followed by a description of the detectors used in our experiments. We visualize our experimental setup with regards to dataset preparation, choice of malware detectors, and the creation of transformed binaries in Figure~\ref{fig:process}.

\subsection{Evaluation Dataset}
\label{sub:eval-dataset}
Our evaluation is based on a sample from the SOREL-20M dataset~\cite{harang2020sorel20m}, a dataset of 20M Windows PE files. This dataset contains hashes of both benign and malicious files, along with their EMBER (LIEF) feature vectors~\cite{ember_dataset}. Each sample is accompanied by metadata, and malicious behavior labels based on details supplied by detection vendors at the point of collection. The \label{test set} is the set of samples on which transformations are to be applied to the original binaries. In our evaluation, we randomly chose 3,362 malware binaries for the test set from the SOREL-20M repository of malware samples. For copyright reasons, SOREL does not distribute benign binaries. As we discuss next, we need to obtain the raw binary to be able to perform transformations. For this reason, the benign hashes obtained from SOREL were queried through the VirusTotal service, and if the sample existed, it was downloaded for evaluation. This resulted in 1,564 test benign binaries.

\subsection{Malware Detectors}
\label{sec:detectors} 
\descr{MalConv.} The MalConv detector from~\cite{raff2018malconv} is our canonical detection model.
This is a neural network architecture designed for malware detection using raw byte sequences of executable files. It is based on a convolutional neural network (CNN) architecture that processes raw byte sequences of executable files and consists of layers including raw byte embedding, 1D convolutions, temporal max-pooling, fully connected layers, and a softmax output layer. The model is designed to consider both local and global contexts while examining an entire file, allowing it to capture important features for malware detection. The model is trained on a dataset of 400,000 executable files split evenly between benign and malicious classes.

\descr{EMBER-feature based RF (EMBER-RF).} We implemented a Random Forest model utilizing the default parameters and trained on the EMBER feature set. The training set for this model consists of 200K balanced malware and Benign samples, subsampled from the SOREL dataset~\cite{harang2020sorel20m}. These training samples are disjoint from the aforementioned evaluation set (c.f. Section~\ref{sub:eval-dataset}). The resulting 
EMBER-RF model achieved an accuracy of 97.09\% when trained and tested on 90\% and 10\% of the dataset respectively. 


\descr{\ssdeep{}.} This is an LSH used widely for signature-based malware detection. An overview of how the algorithm works is shown in Section~\ref{sub:ssdeep}. We take the \ssdeep{} hash of malicious and benign binaries in the test dataset to serve as the local database of the detector, against which any new (transformed) binaries are checked for malicious or benign behavior. 
We use the Python implementation~\cite{ssdeep} 
of \ssdeep{}, which provides a similarity value we shall use to \texttt{compute} a function that gives a similarity score between 0 and 100. This is used as the metric $\mu$ defined in Section~\ref{sub:defs}.

\descr{Ensemble.} We also consider an ensemble detector taking all three detectors into account. The parameter $1 \leq m \leq 3$ determines the decision rule. All decisions are made on the positive label, i.e., malware. The \textit{minority rule}, $m = 1$, means that the program is labelled as malware if at least one detector classifies it as such. Otherwise, it is classified as benign. The \textit{majority rule}, $m = 2$, requires at least two detectors classifying the program as malware. The \textit{consensus rule}, $m = 3$, requires unanimous classification of malware. The choice of $m$ has an obvious impact on the TPR and FPR rates, as we shall experimentally evaluate.

\subsection{Creating Transformed Binaries}
\label{sub:transformed-binaries}
The transformations from malware makeover require the list of functions contained in a program binary. Given a program binary, we therefore (a) disassemble it through IDA~\cite{idapro} to obtain its control flow graph (CFG), and (b) from the CFG, extract functions for the malware makeover algorithm. Once the functions have been extracted we can run the malware makeover algorithm with a target detector. 
Figure \ref{fig:process} depicts our pipeline for creating adversarial binaries. 


For each malware detector, discussed in Section~\ref{sec:detectors}, we formulated a targeted adversarial attack to alter the classification of all binary samples within the evaluation dataset, as depicted in Figure \ref{fig:process}. A whitebox attack strategy was applied to MalConv, leveraging its vulnerability to techniques effective against models based on gradient descent, as discussed in Section~\ref{sec:mmo}. This is the default attack as used by the authors of malware makeover~\cite{lucas2021malware}. In contrast, black-box attacks were employed against EMBER-RF and \ssdeep{}, utilizing differing transformation techniques based on their susceptibility.

For the malware makeover attack on EMBER-RF, we imported the EMBER-RF model into the malware makeover program and executed it. However, to assess whether the modification to the binary is successful, we need to extract EMBER features after each alteration to a binary and evaluate them using the EMBER-RF model. We incorporated the EMBER feature extraction script from \cite{emberpscript, ember_dataset} 
into the malware makeover framework to automate this. However, the integration of the EMBER feature extraction script raised several incompatibility issues, since the EMBER repository's code is in Python 3 and the malware makeover is based on Python 2.7. To address this incompatibility, we initially created a socket to send over the modified binary to the EMBER feature script and get a feature vector. However, this process introduced greater latency. We therefore decided to convert the EMBER feature extraction code to Python 2.7 to ensure compatibility with the malware makeover architecture and make the transformation process faster. 

We also modified the malware makeover's codebase to integrate \ssdeep{}, i.e., computing \ssdeep{} hashes and computing the similarity between the hashes of the original and modified binaries after each alteration until a threshold is reached. Instead of using a modificed attack on \ssdeep{} as outlined in Section~\ref{sub:ssdeep}, we used malware makeover's blackbox algorithm (with the similarity difference used as a proxy for confidence value difference), so that all three attacks can be evaluated under the same framework.  


\section{Transferability of Adversarial Examples}
In this section, we evaluate whether a binary adversarially transformed with a target detector can also induce misclassification in other detectors.

\subsection{Performance of Detectors on Original Dataset.}
Table~\ref{tab:orig} shows the performance of the three detectors on the original binaries in our test dataset (evaluation dataset from Section~\ref{sub:eval-dataset}). Alongside the TPR and FPR, we show the accuracy (Acc), F1 score (FS) and the confusion matrix (CM), which will be used throughout this paper. The performance of all three detectors is very good, with \ssdeep{} showing a 100\% TPR indicating that all binaries were significantly different from one another to cause this hash-based detector to raise any false positives. 

\begin{table}[th]
\renewcommand{\arraystretch}{1.15}
\tabcolsep=0.12cm
\begin{center}
\caption{Performance of malware detectors on the original binaries.} 
\label{tab:orig}
\resizebox{\linewidth}{!}{%
\begin{tabular}{ c | c | c | c | c | c }
 \hline
 & \multicolumn{5}{c}{\bf Performance Metrics}\\
\cline{2-6}
 \textbf{Detector} & \textbf{Acc}&\textbf{FS} &\textbf{TPR} &\textbf{FPR} &\textbf{CM}$\big(\begin{smallmatrix} TN & FP\\ FN & TP \end{smallmatrix}\big)$ \\ [0.5ex]
 \hline\hline
MalConv & 84.34\% & 86.73\% & 80.11\% & 8.18\% & $\begin{smallmatrix}1436 & 128 \\ 550 & 2215\end{smallmatrix}$\\
\hline
EMBER-RF & 81.17\% & 83.95\% & 77.07\% & 11.57\% & $\begin{smallmatrix}1383 & 181 \\ 634 & 2131\end{smallmatrix}$ \\
\hline
SSDeep & 100.00\% & 100.00\% & 100.00\% & 0.00\% & $\begin{smallmatrix}1564 & 0 \\ 0 & 2765\end{smallmatrix}$ \\
\hline
\end{tabular}
}
\end{center}
\end{table}

The table contains only 4,329 out of the 4,926 binaries in our test dataset. This is due to the fact that while running malware makeover with the MalConv detector as the target detector, we could only run 4,329 out of the 4,926 binaries, due to the hard binary size restriction of 2MB in malware makeover, and IDA failing to extract CFGs. Likewise, for various other reasons, malware makeover also caused errors on a number of program binaries with the other two detectors as the target.  

Table \ref{tab:transformation} shows the percentage of malware and benign binaries that were successfully processed by malware makeover given a particular malware detector. We consider a binary as successfully processed if it either evades the target model or malware makeover runs out of transformations to try within a time limit of 3 hours or a maximum number of iterations of 20. The binaries that were not successfully processed are discarded from the pool of binaries for evaluation for that target model. They may or may not have successfully evaded the target model had we given malware makeover more time or more iterations. For EMBER-RF and \ssdeep{}, the percentage of successfully processed binaries is even lower. 

The table also shows the transformation types which were successful in evading a target model. Specifically, the \textit{Disp} transformation was applied to both MalConv and EMBER-RF, while the \textit{IPR} transformation was chosen for its efficacy in circumventing \ssdeep{}. Notably, our findings revealed that the \textit{Disp} transformation, while effective against MalConv and EMBER-RF, did not compromise the integrity of \ssdeep{}, highlighting the unique resilience of each detection system to different adversarial manipulations. This is most likely due to the lack of functions on which \textit{Disp} transformation can be applied even while using higher budgets. 

\begin{table}[hbt!]
\renewcommand{\arraystretch}{1.15}
\tabcolsep=0.12cm
\begin{center}
\caption{Percentage of binaries successfully processed by malware makeover and attack parameters used across malware detectors. \emph{Trans.} and \emph{Iter.} represent the type of transformation and number of iterations, respectively.
} 
\label{tab:transformation}
\resizebox{\linewidth}{!}{%
\begin{tabular}{c | c | c | c | c | c | c}
 \hline
 & \multicolumn{4}{c}{\bf Attack Parameters} & \multicolumn{2}{| c}{\bf  Processed \# (\%)}\\
\cline{2-7}
 \textbf{Target} & \textbf{Type}&\textbf{Trans.} &\textbf{Iter.} &\textbf{Budget} &\textbf{Malware}&\textbf{Benign} \\ [0.5ex]
 \hline\hline
 Total & - & - & - & - & 3362 (100\%) & 1564 (100\%)\\
 \hline
MalConv & whitebox & Disp & 20 & 0.05 & 2765 (82.24\%) & 1564 (100\%)\\
\hline
EMBER-RF & blackbox & Disp & 20 & 0.05 & 1148 (34.15\%) & 212 (13.56\%)\\
\hline
SSDeep & blackbox & IPR & 20 & - & 1889 (56.19\%) & 644 (41.18\%)\\
\hline
\end{tabular}
}
\end{center}
\end{table}

\subsection{Transferability of MalConv Adversarial Examples} 
We first use MalConv as the target detector and use malware makeover's whitebox attack to create adversarial samples of the original binaries. Table~\ref{tab:mc-tgt} shows the resulting performance of the transformed binaries on all three malware detectors. Not surprisingly, the attack is successful on MalConv with a significant decrease in TPR, dropping from 80\% to 38\%, alongside a significant increase in FPR, from 8.18\% to 81.78\%. However, the same transformed binaries have a relatively little impact on the other two detectors. Note that according to our definition of evading detection~\ref{sub:defs}, this shows that the binaries do evade detection against the other two detectors as well, since the FPR has increased together with a decreased TPR. However, relatively speaking the decrease is less prominent. This solidifies our viewpoint that if the target system employs a variety of detectors, adversarial binaries constructed for only one such detector are likely to still be detected by others, and hence these transformed binaries are not highly transferrable. 

\begin{table}[hbt!]
\renewcommand{\arraystretch}{1.15}
\tabcolsep=0.12cm
\begin{center}
\caption{Performance evaluation of {malware detectors at {\bf MalConv} targeted transformed binaries}.}
\label{tab:mc-tgt}
\resizebox{\linewidth}{!}{%
\begin{tabular}{ c | c | c | c | c | c }
 \hline
 & \multicolumn{5}{c}{\bf Performance Metrics}\\
\cline{2-6}
 \textbf{Detector} & \textbf{Acc}&\textbf{FS} &\textbf{TPR} &\textbf{FPR} &\textbf{CM}$\big(\begin{smallmatrix} TN & FP\\ FN & TP \end{smallmatrix}\big)$ \\ [0.5ex]
 \hline\hline
 \multicolumn{6}{c}{Orignal Binaries} \\
 \hline
MalConv & 84.34\% & 86.73\% & 80.11\% & 8.18\% & $\begin{smallmatrix}1436 & 128 \\ 550 & 2215\end{smallmatrix}$\\
\hline
EMBER-RF & 81.17\% & 83.95\% & 77.07\% & 11.57\% & $\begin{smallmatrix}1383 & 181 \\ 634 & 2131\end{smallmatrix}$ \\
\hline
SSDeep & 100.00\% & 100.00\% & 100.00\% & 0.00\% & $\begin{smallmatrix}1564 & 0 \\ 0 & 2765\end{smallmatrix}$ \\
\hline
 \multicolumn{6}{c}{MalConv Targeted Transformed Binaries} \\
 \hline 
MalConv & 31.00\% & 41.44\% & 38.23\% & 81.78\% & $\begin{smallmatrix}285 & 1279 \\ 1708 & 1057\end{smallmatrix}$ \\
 \hline
  EMBER-RF & 77.85\% & 80.98\% & 73.82\% & 15.03\% & $\begin{smallmatrix}1329 & 235 \\ 724 & 2041\end{smallmatrix}$ \\
\hline
SSDeep & 94.48\% & 95.74\% & 97.07\% & 10.10\% & 
$\begin{smallmatrix}1406 & 158 \\ 81 & 2684 \end{smallmatrix}$ \\
\hline
\end{tabular}
}
\end{center}
\end{table}

\begin{figure}[th]
\centering
\includegraphics[width=1\linewidth]{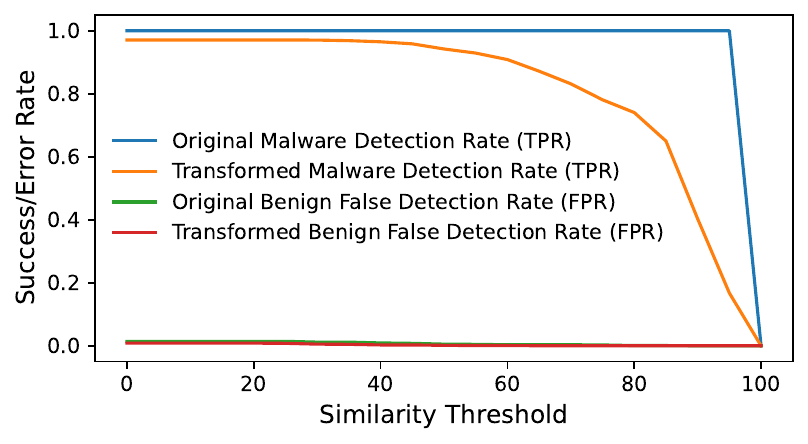}
\caption{\ssdeep{} analysis of original binaries compared with Malconv targeted transformed binaries.} 
\label{fig:ssdeep-malconv}
\end{figure}

For \ssdeep{} we need to determine a similarity (distance) threshold beyond which the hash of a binary is considered not to be similar to a hash in the database. Figure~\ref{fig:ssdeep-malconv} shows the rate of successfully detected a transformed binary in the database through their \ssdeep{} hashes across different similarity thresholds. The rate is relatively unchanged for any threshold less than 0.4. We chose a threshold of 0.25 for \ssdeep{}, which is used for the remainder of this paper.   

\subsection{Transferability of EMBER-RF Adversarial Examples}

To test that non-transferability of the adversarial attack is not just the result of a particular detector, i.e., MalConv, we repeated the above experiment but this time using EMBER-RF as the target model. We use the black-box version of the malware makeover algorithm, since there are no gradients involved in this classifier. Note that the features are now completely changed: raw bytes in MalConv versus EMBER features in EMBER-RF. From Table~\ref{tab:transformation}, we have a reduced set of 1148 malware and 212 binaries for EMBER-RF as the target model, since the rest of the binaries were not successfully processed by malware makeover. We therefore recompute the TPR and FPR of all three detectors on the original binaries from this reduced set shown in Table~\ref{tab:tgt-ember}. For this reduced set EMBER-RF has a fairly high FPR (46.23\%). However, this is further increased and TPR further decreased in the transformed binaries. Thus the changes do successfully evade detection. Once again, these modifications did not signficantly affect the performance of the other detectors; in particular, Malconv's metrics did not change. Thus, these results show that 
adversarial examples that are specially designed to evade EMBER-RF model are not entirely transferable to the other detection methods in our study.



\begin{table}[hbt!]
\renewcommand{\arraystretch}{1.15}
\tabcolsep=0.12cm
\begin{center}
\caption{Performance evaluation of malware detectors at {\bf EMBER-RF} Targeted Transformed Binaries (Black-box).} 
\label{tab:tgt-ember}
\resizebox{\linewidth}{!}{%
\begin{tabular}{ c | c | c | c | c | c }
 \hline
 & \multicolumn{5}{c}{\bf Performance Metrics}\\
\cline{2-6}
 \textbf{Detector} & \textbf{Acc} & \textbf{FS} & \textbf{TPR} &\textbf{FPR} &\textbf{CM}$\big(\begin{smallmatrix} TN & FP\\ FN & TP \end{smallmatrix}\big)$ \\ [0.5ex]
 \hline\hline
 \multicolumn{6}{c}{Orignal Binaries} \\
 \hline
MalConv & 84.41\% & 90.13\% & 84.32\% & 15.09\% & $\begin{smallmatrix}180 & 32 \\ 180 & 968\end{smallmatrix}$\\
\hline
EMBER-RF & 59.12\% & 71.28\% & 60.10\% & 46.23\% & $\begin{smallmatrix}114 & 98 \\ 458 & 690\end{smallmatrix}$ \\
\hline
SSDeep & 100.00\% & 100.00\% & 100.00\% & 0.00\% & $\begin{smallmatrix}212 & 0 \\ 0 & 1148\end{smallmatrix}$ \\
\hline
 \multicolumn{6}{c}{EMBER-RF Targeted Transformed Binaries (Black-box)} \\
 \hline 
MalConv &  84.41\% & 90.13\% & 84.32\% & 15.09\% & $\begin{smallmatrix}180 & 32 \\ 180 & 968\end{smallmatrix}$ \\
\hline
EMBER-RF & 40.00\% & 56.60\% & 46.34\% & 94.34\% & $\begin{smallmatrix}12 & 200 \\ 616 & 532\end{smallmatrix}$\\
\hline
SSDeep & 95.00\% & 97.00\% & 95.82\% & 9.43\% & $\begin{smallmatrix}192 & 20 \\ 48 & 1100\end{smallmatrix}$ \\
  \hline
  \hline
\end{tabular}
}
\end{center}
\end{table}


\subsection{Transferability of \ssdeep{} Adversarial Examples}
Following the evaluation of malware detectors on MalConv and Ember-RF adversarial examples, we first establish baseline performance by only selecting those original samples that were successfully processed by the malware makeover algorithm for \ssdeep{}: 1889 malware and 644 benign, from Table~\ref{tab:transformation}. Table~\ref{tab:tgt-sd} presents the performance of the three malware detectors against \ssdeep{} targeted transformed binaries. There is a sharp rise in FPR from 0\% to 96.27\% and a substantial drop in TPR from 100\% to 16.52\% for \ssdeep{}. Once again, the impact on EMBER-RF is minimal, while MalConv's metrics showed improvement. These results indicate that adversarial examples crafted to circumvent \ssdeep detection capabilities fail to affect the efficacy of other malware detectors significantly.


\begin{table}[hbt!]
\renewcommand{\arraystretch}{1.15}
\tabcolsep=0.12cm
\begin{center}
\caption{Performance evaluation of {\it Malware detectors at {\bf SSDEEP} Targeted Transformed Binaries (Black-box)}}. 
\label{tab:tgt-sd}
\resizebox{\linewidth}{!}{%
\begin{tabular}{ c | c | c | c | c | c }
 \hline
 & \multicolumn{5}{c}{\bf Performance Metrics}\\
\cline{2-6}
 \textbf{Detector} & \textbf{Acc}&\textbf{FS} &\textbf{TPR} &\textbf{FPR} &\textbf{CM}$\big(\begin{smallmatrix} TN & FP\\ FN & TP \end{smallmatrix}\big)$ \\ [0.5ex]
 \hline\hline
 \multicolumn{6}{c}{Orignal Binaries} \\
 \hline
MalConv & 84.64\% & 88.68\% & 80.68\% & 3.73\% & $\begin{smallmatrix}620 & 24 \\ 365 & 1524\end{smallmatrix}$ \\
\hline
EMBER-RF & 86.06\% & 90.07\% & 84.75\% & 10.09\% & $\begin{smallmatrix}579 & 65 \\ 288 & 1601\end{smallmatrix}$ \\
\hline
SSDeep & 100.00\% & 100.00\% & 100.00\% & 0.00\% & $\begin{smallmatrix}644 & 0 \\ 0 & 1889\end{smallmatrix}$ \\
\hline
 \multicolumn{6}{c}{SSDEEP Targeted Transformed Binaries} \\
 \hline 
MalConv &  85.08\% & 89.01\% & 81.00\% & 2.95\% & $\begin{smallmatrix}625 & 19 \\ 359 & 1530\end{smallmatrix}$ \\
\hline
EMBER-RF & 76.94\% & 82.87\% & 74.80\% & 16.77\% & $\begin{smallmatrix}536 & 108 \\ 476 & 1413\end{smallmatrix}$ \\
\hline
SSDeep & 13.26\% & 22.12\% & 16.52\% & 96.27\% & 
$\begin{smallmatrix}24 & 620 \\ 1577 & 312\end{smallmatrix}$\\
  \hline
  \hline
\end{tabular}
}
\end{center}
\end{table}


\section{Performance of the Ensemble Detector}
\label{sec:ensemble}
Our results from previous section indicate that a possible mitigation technique against adversarially transformed binaries is to use an ensemble of detectors. In this section we this ensemble approach, incorporating all three detectors—MalConv, EMBER-RF, and \ssdeep{}—to determine the efficacy of an ensemble in detecting transformed binaries. This analysis involves selecting a subset of adversarial examples that have been successfully processed by the malware makeover algorithm for all three detectors. For the purpose of establishing a baseline performance, a corresponding subset of original binaries is also compiled. The performance of each detector, providing individual verdicts/classification on this common subset of original and transformed binaries, is presented in Table \ref{tab:perform-ml-individual}. A key thing to note here is that unfortunately the performance of EMBER-RF on the original binaries in this common subset is lower than its performance on the complete list of original binaries (see Table~\ref{tab:orig}). This discrepancy has been factored in our conclusions in the following.

\begin{table}[hbt!]
\renewcommand{\arraystretch}{1.15}
\tabcolsep=0.12cm
\begin{center}
\caption{Performance evaluation of {\it {individual} malware detectors on the common test set}.} 
\label{tab:perform-ml-individual}
\resizebox{\linewidth}{!}{
\begin{tabular}{ c | c | c | c | c | c }
 \hline
 & \multicolumn{5}{c}{\bf Performance Metrics}\\
\cline{2-6}
 \textbf{Detector} & \textbf{Acc}&\textbf{FS} &\textbf{TPR} &\textbf{FPR} &\textbf{CM}$\big(\begin{smallmatrix} TN & FP\\ FN & TP \end{smallmatrix}\big)$ \\ [0.5ex]
 \hline\hline
 \multicolumn{6}{c}{Orignal Binaries} \\
 \hline
MalConv & 78.64\% & 81.64\% & 75.74\% & 3.96\% & $\begin{smallmatrix}97 & 4 \\ 147 & 459\end{smallmatrix}$ \\
\hline
EMBER-RF & 71.43\% & 75.28\% & 72.77\% & 36.63\% & $\begin{smallmatrix}64 & 37 \\ 165 & 441\end{smallmatrix}$ \\
\hline
SSDeep & 100.00\% & 100.00\% & 100.00\% & 0.00\% & $\begin{smallmatrix}101 & 0 \\ 0 & 606\end{smallmatrix}$ \\
\hline
 \multicolumn{6}{c}{MalConv Targeted Transformed Binaries} \\
 \hline 
MalConv & 37.91\% & 46.42\% & 39.77\% & 73.27\% & $\begin{smallmatrix}27 & 74 \\ 365 & 241\end{smallmatrix}$ \\
\hline
EMBER-RF & 56.15\% & 62.42\% & 61.88\% & 78.22\% & $\begin{smallmatrix}22 & 79 \\ 231 & 375\end{smallmatrix}$ \\
\hline
SSDeep & 95.62\% & 97.44\% & 97.19\% & 13.86\% & $\begin{smallmatrix}87 & 14 \\ 17 & 589\end{smallmatrix}$ \\
\hline
 \multicolumn{6}{c}{Ember-RF (BB) Targeted Transformed Binaries} \\
 \hline 
MalConv & 89.39\% & 90.27\% & 89.27\% & 9.90\% & $\begin{smallmatrix}91 & 10 \\ 65 & 541\end{smallmatrix}$ \\
\hline
EMBER-RF & 53.04\% & 59.59\% & 61.06\% & 95.05\% & $\begin{smallmatrix}5 & 96 \\ 236 & 370\end{smallmatrix}$ \\
\hline
SSDeep & 95.90\% & 97.61\% & 97.52\% & 13.86\% & $\begin{smallmatrix}87 & 14 \\ 15 & 591\end{smallmatrix}$ \\
\hline
 \multicolumn{6}{c}{SSDeep Targeted Transformed Binaries} \\
 \hline 
MalConv & 78.93\% & 81.86\% & 76.24\% & 4.95\% & $\begin{smallmatrix}96 & 5 \\ 144 & 462\end{smallmatrix}$ \\
\hline
EMBER-RF & 65.49\% & 70.52\% & 66.34\% & 39.60\% & $\begin{smallmatrix}61 & 40 \\ 204 & 402\end{smallmatrix}$ \\
\hline
SSDeep & 8.91\% & 16.36\% & 10.40\% & 100.00\% & $\begin{smallmatrix}0 & 101 \\ 543 & 63\end{smallmatrix}$ \\
  \hline
  \hline
\end{tabular}
}
\end{center}
\end{table}

\descr{Minority Rule.} We start with the minority rule ($m=1$) as defined in Section~\ref{sec:detectors}. 
Table~\ref{tab:perform-ml-OR} shows the performance evaluation of the minority rule ensemble on both original and transformed common subset. The ensemble resulted in a perfect TPR of 100\% on original binaries, indicating that all malicious binaries were correctly identified. However, the 39.60\% FPR suggests that a significant portion of benign binaries were incorrectly classified as malicious. This is mainly due to EMBER-RF's poor FPR on the original binaries. On the transformed binaries, while the TPR remains high (>98\%) for all adversarial sets, there is a sharp increase in FPR to 95.05\% and 96.04\% in case of MalConv and EMBER-RF adversarial set showing that the ensemble incorrectly classifies a large number of benign binaries as malicious. The minority rule however performs comparatively better on the \ssdeep{} adversarial set, with a high TPR of 99.50\% and a slight increase in FPR to 46.53\% when compared to the original set, indicating a better balance between identifying malicious and benign binaries compared to the other transformed sets. The high FPR of the minority rule ensemble is not surprising, as only a single detector can influence the decision on a program being a malware, which includes the target model as well. 



\begin{table}[hbt!]
\renewcommand{\arraystretch}{1.15}
\tabcolsep=0.12cm
\begin{center}
\caption{Performance evaluation of {{\bf Minority Rule} ensemble on original and transformed binaries}. TTB stands for targetted transformed binaries.} 
\label{tab:perform-ml-OR}
\resizebox{\linewidth}{!}{%
\begin{tabular}{c | c | c | c | c | c}
 \hline
 & \multicolumn{5}{c}{\bf Performance Metrics}\\
\cline{2-6}
 \textbf{Binaries} & \textbf{Acc}&\textbf{FS} &\textbf{TPR} &\textbf{FPR} &\textbf{CM}$\big(\begin{smallmatrix} TN & FP\\ FN & TP \end{smallmatrix}\big)$ \\ [0.5ex]
 \hline\hline
Orignal Binaries & 94.34\% & 96.81\% & 100.00\% & 39.60\% & $\begin{smallmatrix}61 & 40 \\ 0 & 606\end{smallmatrix}$ \\
\hline
 MalConv TTB & 85.29\% & 92.00\% & 98.68\% & 95.05\% & $\begin{smallmatrix}5 & 96 \\ 8 & 598\end{smallmatrix}$ \\
\hline
EMBER-RF TTB & 86.14\% & 92.51\% & 99.83\% & 96.04\% & $\begin{smallmatrix}4 & 97 \\ 1 & 605\end{smallmatrix}$\\
\hline
SSDeep TTB & 92.93\% & 96.02\% & 99.50\% & 46.53\% & $\begin{smallmatrix}54 & 47 \\ 3 & 603\end{smallmatrix}$ \\
\hline
\end{tabular}
}
\end{center}
\end{table}

\descr{Majority Rule.} From Table \ref{tab:perform-ml-maj}, it can be seen that the majority rule ensemble achieves a high TPR of 94.39\% with an extremely low FPR of 0.99\%, indicating excellent detection of actual malicious binaries and rarely misclassifying benign files as malicious. However, there's a noticeable decline in TPR to 76.40\% and a considerable rise in FPR to 63.37\%, suggesting that the ensemble is less effective at detecting both malicious and benign binaries when dealing with MalConv-evasive binaries. This again is mostly due to EMBER-RF's poor performance on these transformed binaries (see Table~\ref{tab:perform-ml-individual}). This explains why the ensemble achieves a high TPR (>92\%) and a reasonably low FPR (<21\%) when EMBER-RF is the target model, as the majority now includes the other two detectors, MalConv and \ssdeep{} which show good TPR and FPR. The performance of the marjority rule is also low when \ssdeep{} is the target, once again owing mostly to EMBER-RF's poor FPR on the original samples. From these observations, the majority rule ensemble is effective in still producing a high TPR and low FPR as long as the majority of the detectors are not evaded by a transformation technique. 


\begin{table}[hbt!]
\renewcommand{\arraystretch}{1.15}
\tabcolsep=0.12cm
\begin{center}
\caption{Performance evaluation of {{\bf Majority Rule} ensemble at original and transformed binaries}. TTB stands for targetted transformed binaries.} 
\label{tab:perform-ml-maj}
\resizebox{\linewidth}{!}{%
\begin{tabular}{c | c | c | c | c | c}
 \hline
 & \multicolumn{5}{c}{\bf Performance Metrics}\\
\cline{2-6}
 \textbf{Binaries} & \textbf{Acc}&\textbf{FS} &\textbf{TPR} &\textbf{FPR} &\textbf{CM}$\big(\begin{smallmatrix} TN & FP\\ FN & TP \end{smallmatrix}\big)$ \\ [0.5ex]
 \hline\hline
Original & 95.05\% & 97.03\% & 94.39\% & 0.99\% & $\begin{smallmatrix}100 & 1 \\ 34 & 572\end{smallmatrix}$\\
\hline
 MalConv TTB & 70.72\% & 81.73\% & 76.40\% & 63.37\% & $\begin{smallmatrix}37 & 64 \\ 143 & 463\end{smallmatrix}$\\
\hline
EMBER-RF TTB & 92.50\% & 95.59\% & 94.72\% & 20.79\% & $\begin{smallmatrix}80 & 21 \\ 32 & 574\end{smallmatrix}$\\
\hline
SSDeep TTB & 54.74\% & 67.28\% & 54.29\% & 42.57\% & $\begin{smallmatrix}58 & 43 \\ 277 & 329\end{smallmatrix}$\\
\hline
\end{tabular}
}
\end{center}
\end{table}

\descr{Consensus Rule.} 
The consensus rule ensemble requires all detectors to agree that a sample is malicious before classifying it as such. This strategy tends to yield a lower TPR but a much lower FPR, aiming for precision in identifying malware. The results for consensus rule are presented in Table~\ref{tab:perform-ml-AND}. On the original binaries, the FPR is at 0\%, indicating no benign files are misclassified as malicious. The TPR is moderate at 54.13\%, indicating that only over half of the actual malware are detected. The TPR significantly drops to 23.76\%, showing that the ensemble misses the majority of malware transformed to evade MalConv; however, there is a slight decrease in TPR to 53.30\% and 48.18\% (when compared to the original binaries) in the Ember-RF and SSDeep adversarial examples, suggesting that these transformations slightly affect the consensus rule malicious detection capability. On the other hand, the FPR increases modestly to 6.93\% for the MalConv adversarial set and slightly increases to 1.98\% and 0.99\% for the Ember-RF and SSDeep adversarial sets, respectively, which is still relatively low, suggesting that few benign files are misclassified as malicious.

Keeping this in mind, the consensus rule ensemble appears to err on the side of caution, opting for high specificity (low FPR) at the expense of sensitivity (TPR). This means it is highly reliable at confirming benign files but at the risk of failing to detect a significant amount of actual malware, particularly when dealing with adversarially transformed malware targeting MalConv.

\begin{table}[hbt!]
\renewcommand{\arraystretch}{1.15}
\tabcolsep=0.12cm
\begin{center}
\caption{Performance evaluation of {{\bf Consensus Rule} ensemble on original and transformed binaries}. TTB stands for targetted transformed binaries.} 
\label{tab:perform-ml-AND}
\resizebox{\linewidth}{!}{%
\begin{tabular}{c | c | c | c | c | c}
 \hline
 & \multicolumn{5}{c}{\bf Performance Metrics}\\
\cline{2-6}
 \textbf{Binaries} & \textbf{Acc}&\textbf{FS} &\textbf{TPR} &\textbf{FPR} &\textbf{CM}$\big(\begin{smallmatrix} TN & FP\\ FN & TP \end{smallmatrix}\big)$ \\ [0.5ex]
 \hline\hline
Original & 60.68\% & 70.24\% & 54.13\% & 0.00\% & $\begin{smallmatrix}101 & 0 \\ 278 & 328\end{smallmatrix}$\\
\hline
MalConv TTB & 33.66\% & 38.04\% & 23.76\% & 6.93\% & $\begin{smallmatrix}94 & 7 \\ 462 & 144\end{smallmatrix}$\\
\hline
EMBER-RF TTB & 59.69\% & 69.39\% & 53.30\% & 1.98\% & $\begin{smallmatrix}99 & 2 \\ 283 & 323\end{smallmatrix}$\\
\hline
SSDeep TTB & 55.45\% & 64.96\% & 48.18\% & 0.99\% & $\begin{smallmatrix}100 & 1 \\ 314 & 292\end{smallmatrix}$\\
\hline
\end{tabular}
}
\end{center}
\end{table}

\section{Are Highly Transformed Binaries Detectable?}
We return to our discussion in the introduction on the necessity of keeping perturbations to a program binary at a minimum. Recall that unlike the image domain, an adversarially transformed program binary need not look similar from a static code analysis point of view to its original binary, as long as it maintains functionality. However, it has been pointed out in the literature that too many changes to a program, e.g., via obfuscation, would make it easy to detect a transformed binary as it would create new signatures~\cite[\S 4]{anderson2018learning}. We first analyze how much of the binaries has been transformed against each target detector, and then use a machine learning algorithm to distinguish between original and transformed binaries.

\subsection{Amount of Adversarial Perturbation} 
\label{sub:levenshtein}

\begin{figure}[th]
\centering
\includegraphics[width=1\linewidth]{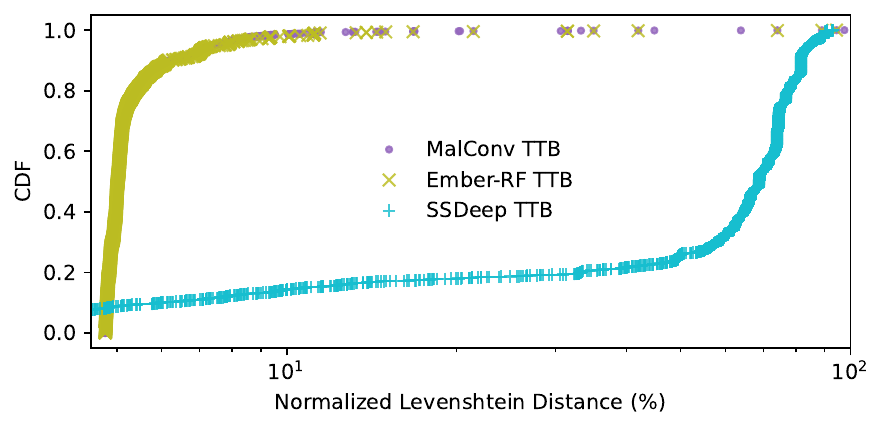}
\caption{Cumulative Distribution Function (CDF) vs. Normalized Levenshtein Distance (NLD) for All Three Adversarial Transformations}
\label{fig:ld_cdf_combine}
\end{figure}

Recall that one of the goals of the attacker is to keep the amount of transformations to a minimum. To check how much perturbation is applied to a binary, we use the normalized Levenshtein distance (NLD) on the original and transformed binaries for each of the test sets used in Table~\ref{tab:mc-tgt} (MalConv), \ref{tab:tgt-ember} (EMBER-RF) and \ref{tab:tgt-sd} (\ssdeep{}). We show the cumulative distribtuion function (CDF) of the NLD between the original and tranformed binaries in Figure~\ref{fig:ld_cdf_combine}.  Majority of the binaries transformed with MalConv and EMBER-RF as targets, are within 10\% NLD of their original counterparts, indicating little changes to the codes of these binaries. On the other hand, the binaries transformed for \ssdeep{} evasion are substantially changed. Note that this is also due to the different transformation technique used for this detector: IPR vs Disp. From the adversarial point-of-view, too many changes are not ideal as we discuss in the next section.



\subsection{Detecting Highly Transformed Binaries}
\label{sub:highly-transformed}

 To evaluate whether highly transformed binaries are more easily detectable we trained two machine learning models to classify original and transformed binaries, i.e., transformed via malware makeover. The first classifier is a random forest (RF) model while the second is the LightGBM model (briefly discussed in \S~\ref{sec:detectors}), which is a tree-based lightweight gradient boosting framework built on an ensemble of weak learners (typically decision trees) to create a strong predictive model.  
 The two models were trained on a data set split into 80\% for training and 20\% for testing, which included both original and transformed binaries. For both these models we used raw bytes as the features. The results in Section~\ref{sub:levenshtein} show that when the target model is MalConv or EMBER-RF the binaries are only slightly transformed in terms of the normalized Levenshtein distance (NLD). On the other hand, the binaries are heavily transformed when the target is \ssdeep{} again with respect to NLD. Thus, we would expect the models to show better performance on the latter.

Our results summarized in Table~\ref{tab:detect-trans} are in contrast to this impression.  The performance for MalConv and EMBER-RF targetted transformed binaries is comparable with TPRs of 61.37\% and 58.18\% and FPRs of 35.07\% and 31.97\%, respectively, indicating moderate success. But the RF model shows improved performance on binaries transformed via \ssdeep{}: 70.36\% TPR and 27.51\% FPR. However, the LightGBM model shows an almost perfect TPR of more than 99\% and relatively low FPR of 0-2\%, while showing similar performance to the RF model on \ssdeep{} transformed binaries. This discrepancy may be mostly due to the difference between the types of transformation used in the three detectors: MalConv and EMBER-RF use the Disp transformations whereas \ssdeep{} uses the IPR transformations. 

We hypothesize that the high classification performance of detecting Disp transformations, is the inclusion of semantic {\tt nop} operations. These semantic {\tt nops} may present as unusual, out-of-place values of high entropy which are leveraged as an indicator for transformation by the classifiers of this section. On the other hand, \ssdeep{}'s IPR operations are more subtle, thereby blending into the original binaries, however, with too many replacements, as seen in the binaries of large NLD, detection becomes possible again.

Therefore to test whether high transformations lead to better detection, we trained the LightGBM model on \ssdeep{} original and transformed binaries only, as we focus on only one type of transformation. We then test the models performance on the top 20 and bottom 20 transformed binaries in terms of NLD. Our results in Table~\ref{tab:extreme-binaries} show that the model almost perfectly predicts highly transformed binaries, and shows no better than random guess on binaries that are least transformed. This provides evidence that highly transformed binaries may be easier to detect as malware, as one would expect benign binaries to not be using malware makeover or other transformation techniques so heavily. This justifies the adversarial objective of making minimum perturbations to the original program.

\begin{table}[hbt!]
\renewcommand{\arraystretch}{1.15}
\tabcolsep=0.12cm
\begin{center}
\caption{Comparative performance of the {{\bf Random Forest} and {{\bf LightGBM} models on detecting adversarially transformed binaries with MalConv, Ember-RF, and \ssdeep{} as targets.} TTB stands for targetted transformed binaries.}} 
\label{tab:detect-trans}
\resizebox{\linewidth}{!}{%
\begin{tabular}{c | c | c | c | c | c}
 \hline
 & \multicolumn{5}{c}{\bf Performance Metrics}\\
\cline{2-6}
 \textbf{Binaries} & \textbf{Acc}&\textbf{FS} &\textbf{TPR} &\textbf{FPR} &\textbf{CM}$\big(\begin{smallmatrix} TN & FP\\ FN & TP \end{smallmatrix}\big)$ \\ [0.5ex]
 \hline\hline
 \multicolumn{6}{c}{{\bf Random Forest Model}} \\
 \hline 
 MalConv TTB & 63.11\% & 63.04\% & 61.37\% & 35.07\% & $\begin{smallmatrix}548 & 296 \\ 343 & 545\end{smallmatrix}$\\
\hline
EMBER-RF TTB & 63.05\% & 61.42\% & 58.18\% & 31.97\% & $\begin{smallmatrix}183 & 86 \\ 115 & 160\end{smallmatrix}$\\
\hline
SSDeep TTB & 71.45\% & 70.62\% & 70.36\% & 27.51\% & $\begin{smallmatrix}419 & 159 \\ 163 & 387\end{smallmatrix}$\\
\hline
 \multicolumn{6}{c}{{\bf LightGBM Model}} \\
 \hline 
MalConv TTB & 99.65\% & 99.66\% & 99.77\% & 0.47\% & $\begin{smallmatrix}840 & 4 \\ 2 & 886\end{smallmatrix}$\\
\hline
EMBER-RF & 98.53\% & 98.56\% & 99.27\% & 2.23\% & $\begin{smallmatrix}263 & 6 \\ 2 & 273\end{smallmatrix}$\\
\hline
SSDeep & 70.81\% & 71.21\% & 71.35\% & 29.75\% & $\begin{smallmatrix}366 & 155 \\ 153 & 381\end{smallmatrix}$\\
\hline
\end{tabular}
}
\end{center}
\end{table}

\begin{table}[hbt!]
\renewcommand{\arraystretch}{1.15}
\tabcolsep=0.12cm
\begin{center}
\caption{Comparative performance of the {{\bf LightGBM} model on highly modified (top 20) and least modified (bottom 20) \ssdeep{} targeted transformed binaries with respect to NLD.}}
\label{tab:extreme-binaries}
\resizebox{\linewidth}{!}{%
\begin{tabular}{c | c | c | c | c | c}
 \hline
 & \multicolumn{5}{c}{\bf Performance Metrics}\\
\cline{2-6}
 \textbf{Binaries} & \textbf{Acc}&\textbf{FS} &\textbf{TPR} &\textbf{FPR} &\textbf{CM}$\big(\begin{smallmatrix} TN & FP\\ FN & TP \end{smallmatrix}\big)$ \\ [0.5ex]
 \hline\hline
SSDeep-Top-20 & 97.5\% & 97.56\% & 100\% & 5\% & $\begin{smallmatrix}19 & 1 \\ 0 & 20\end{smallmatrix}$\\
 \hline 
SSDeep-Bottom-20 & 52.5\% & 53.66\% & 55\% & 50\% & $\begin{smallmatrix}10 & 10 \\ 9 & 11\end{smallmatrix}$\\

\hline
\end{tabular}
}
\end{center}
\end{table}

\section{Results on VirusTotal} 

\begin{figure}[!t]
\centering
\begin{subfigure}{\linewidth}
    \centering
    \includegraphics[width=1\textwidth]{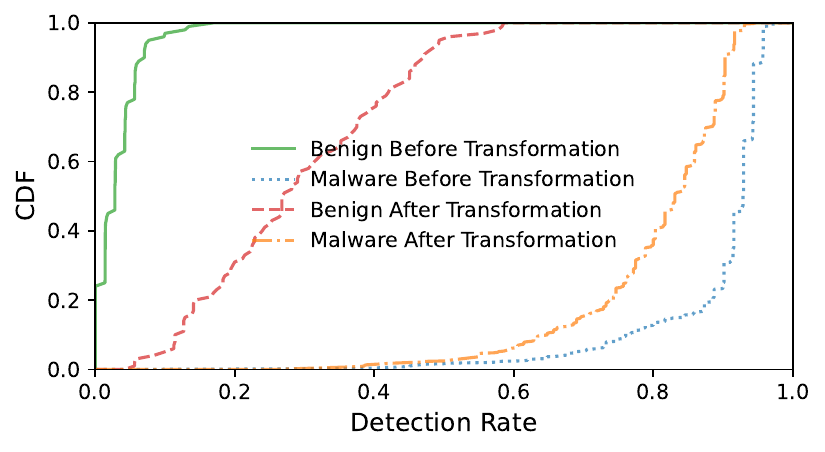}
    \caption{{\bf MalConv} transformation.}
    \label{fig:mc-vt}
\end{subfigure}
\begin{subfigure}{\linewidth}
    \centering
    \includegraphics[width=1\textwidth]{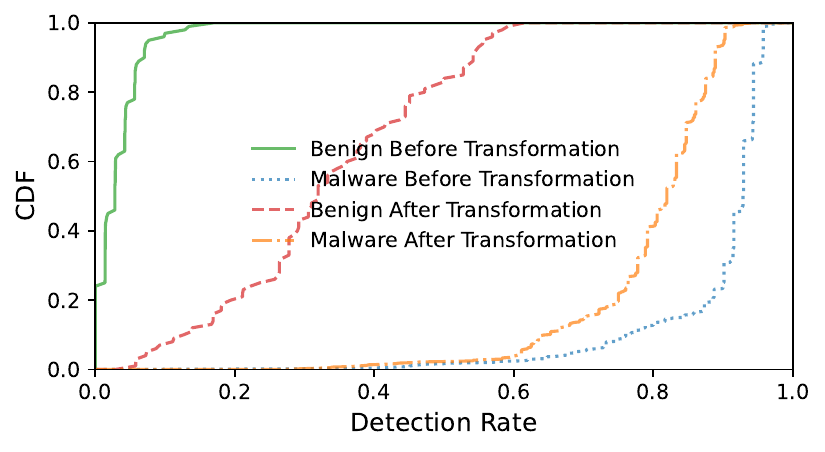}
    \caption{{\bf Ember-RF} transformation.}
    \label{fig:bb-vt}
\end{subfigure}
\begin{subfigure}{\linewidth}
    \centering
    \includegraphics[width=1\textwidth]{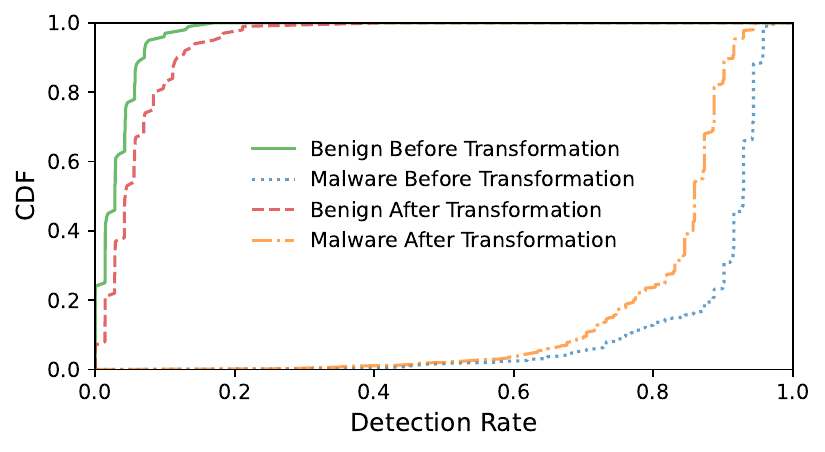}
    \caption{{\bf SSDeep} transformation.}
    \label{fig:ssdeep-vt}
\end{subfigure}
\caption{CDFs of VirusTotal detection rates for benign and malware {\it before} and {\it after} transformation.}
\label{fig:vt_transformation}
\end{figure}

A major motivation for us to assess the performance of an ensemble of detectors to mitigate adversarial examples is the use of reports from multiple security vendors (scanners) by VirusTotal. This provides a buffer against misclassifications as long as there is consensus amongst multiple scanners. To check the detection rate of the transformed binaries by VirusTotal, we took the 101 benign and 606 malware binaries (total of 707 binaries) common amongst the three target detectors (Table~\ref{tab:perform-ml-individual}). We then use the VirusTotal API with each original binary and its transformed variant to obtain labels returned by the VirusTotal scanners. The total number of scanners vary across binaries, with a minimum of 51 and a maximum of 72 scanners, averaging 70.78\% per binary. From the labels thus returned, we define the \textit{detection rate} as being the fraction of scanners labelling the given binary as malware.   

Figures~\ref{fig:vt_transformation} (a), (b) and (c) show results for binaries when the target detectors are MalConv, EMBER-RF and \ssdeep{}, respectively. For all three detectors, the detection rate of transformed malware binaries shows a marked decrease. On the other hand, the detection rate increases for transformed benign binaries, more prominently for MalConv and EMBER-RF, indicating that the  transformation techniques fool the scanners into thinking that the given binary is a malware. The transformed binaries with \ssdeep{} as the target detector were less likely to be misclassified than the other two detectors. If we use the majority rule, i.e., detection rate of 0.5, this indicates that very few malware binaries would be misclassified by the scanners, indicating that an effective measure to prevent adversarial attacks on malware binaries is to use an ensemble of detectors with majority rule, consistent with our analysis in Section~\ref{sec:ensemble}. Furthermore, if the adversary tries to evade multiple detectors this will inevitably increase the amount of perturbation in the original binary resulting in the binary being detected as malware as discussed in Section~\ref{sub:highly-transformed}.





\section{Related Work}

Our work investigates the robustness of malware detectors against adversarial software binaries when transferred to detectors of different feature sets and model architectures. There are many related works in the space of detecting malicious software binaries, we take a moment to position this working within this broad field.

\descr{Adversarial Examples.}
Adversarial malware borrows conventions from the genesis of adversarial examples, an area whereby minor imperceptible modifications are made to the input domain, typically on numeric vectors, and images, to induce a misclassification on a machine learning model~\cite{carlini2017towards}. Deep learning models, the architecture that has found increasing employ in malware detection, is particularly vulnerable to such adversarial inputs.
The definition provided in adversarial examples for a need to constrain the size of perturbation apply to adversarial malware~\cite{kurakin2018adversarial,yuan2019adversarial}, as evidenced in Section~\ref{sub:levenshtein}, whereby SSDeep transformed binaries are more easily detected whence larger transformations from their original binary are made. The transformation techniques within this paper are not optimized to minimize the size of their perturbations, they are instead only able to be capped within a specified budget. There remains open work for the application of these functionality preserving techniques to holistically optimize transformations across the binary.

\descr{Deep Malware Classification.}
Deep learning techniques have enabled the rapid analysis of malware, with works to directly ingest binary data for classification, or with features extracted prior to processing~\cite{kalash2018malware}.
Raw-binary classifiers take as input the bytes of the binary directly, with the feature abstraction and optimization left solely to a machine learning model~\cite{raff2018malconv}, or processed directly as a hash (e.g. \ssdeep{})~\cite{ssdeep}.
On the other hand if features are to be extracted, either Static or Dynamic analysis is applied. Static analysis dives into the binary without executing the binary, often, the binary is decompiled\cite{ghidra,idapro} from which features can be extracted, (e.g. EMBER~\cite{ember_dataset}).
Dynamic executes the binaries to observe program behavior to determine maliciousness~\cite{or2019dynamic}.

\descr{Adversarial Training.}
The proliferation of Deep Learning models within malware detection, has provided avenues for the discovery of adversarial malware examples, however the defensive action has not been still, with adversarial training strategies incorporated into DL training.
For example, MalConv~\cite{raff2018malconv}, a raw-binary deep classification model can be adversarially trained on binaries to improve it's robustness against adversarial binaries produced by Malware Makeover~\cite{lucas2023adversarial}. A key insight is that while the adversary is constrained to producing functional binaries, the defender in creating adversarial software binaries is not. As such the range of transformations available to the defender's training set of adversarial examples can be produced much faster, and in greater quantities than the attacker, thereby providing samples for a more adversarially robust model~\cite{zhang2021enhanced}.

\descr{Hash-based detection.}
Our use of \ssdeep{} departs from the initial targets of adversarial malware creation, but nevertheless, is a representative example of similarity-based detection.
Marcelli et al \cite{marcelli-ml-binary} present a comprehensive survey on binary program similarity, and it's varying nuances in detection performance when applied on different computing platforms (architectures), However, they do not consider the robustness of these similarity detectors in the adversarial setting just as we do.
Our evaluation includes obtaining reports from VirusTotal's scanner aggregation service, while these metrics are related to anti-virus scanners, VirusTotal additionally provides numerous similarity based hashes as part of it's analysis. VirusTotal's hashes include Block-Based Hashing (e.g. dcfldd),
Context-Triggered Piece-wise Hashing (e.g. ssdeep),
Statistically-Improbable Features (e.g. sdhash),
Block-Based Rebuilding (e.g. mvHash-B)\footnote{\url{https://www.gdatasoftware.com/blog/2021/09/an-overview-of-malware-hashing-algorithms}}. Given the adoption of these hashing techniques, works have set our to empirically measure their performance in detection~\cite{pagani-beyond, marcelli-ml-binary}, and clustering to discover malware families~\cite{fuzzy-mal-cluster}.

\descr{Union of Deep Learning and Hashing.}
DeepHashing is a domain at the intersection of deep learning techniques, and the typically cryptography grounded space of hashing. NeuralHash~\cite{struppek2022neuralhash} is one such example of \emph{deep hashing}, a combination of a deep neural network (to extract features) and then a locality sensitive hashing technique to ``bin'' similar feature vectors.

\descr{Transferability.} Ours is not the first work to discuss transferability of adversarial attacks in malware detection. The work in~\cite{demontis2019adversarial} discusses transferability of attacks on malware samples from the DREBIN dataset using machine learning models trained using gradient descent. However, they use the same feature space for all models. Furthermore, the transferability rate is lower for programs versus images (respectively, Figures 12 and 7 in~\cite{demontis2019adversarial}). The work in~\cite{pierazzi2020problemspace} shows a \emph{problem-space} attack on Android malware that adds benign code to malware to induce misclassification. The attack is demonstrated on SVM-based classifiers. We note that since these transformations are additions to the malware code, they keep the original malware code, and hence the transformation technique is likely to be detected by \ssdeep{}-based detection or via raw byte sequences. The authors also note that adding heavily obfuscated code may be easier to detect; a point that we have already discussed in the introduction and experimentally verified in Section~\ref{sub:highly-transformed}. Hu et al~\cite{hu2022transferability} demonstrate transferability of adversarial attacks on malware detection across a diverse range of models including neural networks, SVM, logistic regression, and random forest. However, they use the same feature space across each model. In our case, one of the defining factors of the relative lack of transferability is that the feature spaces across the three detectors is different.



\section{Conclusion and Future Work}
We have shown that a malware binary adversarially transformed to evade a target malware detector is not in general successful in evading multiple detectors, especially if those detectors are architecturally different. We show this using a detector based on raw bytes of the binary, another which extracts (EMBER) features from the binary, and a third detector based on changes on disjoint chunks of a binary (locality-sensitive hashing). A straightforward conclusion to mitigate the impact of adversarially transformed binaries is to use an ensemble of detectors with majority rule, as we have discussed. While the adversary can still evade detection against this ensemble by transforming the binary to evade a majority of detectors, this increases the amount of perturbation to the binary resulting in unique signatures being created for the transformation techniques. As we have shown, highly transformed binaries are easily detectable. An interesting area for future work is to develop an adversarial transformation algorithm that can defeat a majority of detectors in an ensemble while keeping changes to a minimum. 

\section*{Acknowledgements}
This work was partially supported by the Australian Defence Science and Technology (DST) Group under the Next Generation Technology Fund (NGTF) scheme.

\bibliographystyle{ieeetr}
\bibliography{robust-bcs.bib}

\appendix
\section*{Appendix}

\section{EMBER Features}
\label{app:ember_feat}

\descr{Parsed features.} LIEF ~\cite{LIEF} facilitates the identification of symbolic strings, characteristics, and properties of the PE file under analysis to extract following meaningful features:
\begin{itemize}[noitemsep,topsep=0pt,parsep=0pt,partopsep=0pt, leftmargin=15pt]
  \item \textit{General file information.} Metrics including file size, details on imports, exports, debug elements, resources, relocations, and symbols derived from the PE header.
  \item \textit{Header information.} Metrics including file timestamps, target machine specifications, image characteristics,  DLL characteristics, file magic from COFF and optional headers.
  \item \textit{Imported functions.} Sourced from the import address table to list functions imported through libraries, achieved through a hashing representation to uniquely represent libraries and functions.
  \item \textit{Exported functions.} Details on exported functions, that utilize a 128-bin hash for the extraction of model features.
  \item \textit{Section information.} Insights into section properties including name, size, entropy, virtual size, and characteristics, leveraging the same hashing trick for detailing section and entry point features.
  \item \textit{DataDirectories} Sourced from the optional header, act as references to subsequent sections, encompassing tables for exports, imports, resources, exceptions, debug data, certificates, and relocation tables.
\end{itemize}

\descr{Format-Agnostic Features.} Beyond parsed elements, the EMBER feature set incorporates three types of format-agnostic features, which do not necessitate PE file parsing:

\begin{itemize}[noitemsep,topsep=0pt,parsep=0pt,partopsep=0pt,leftmargin=15pt]
  \item \textit{Byte histogram.} A 256-value integer array denoting byte normalized frequencies within the file.
  \item \textit{Byte-entropy histogram.} Approximates joint entropy and byte distributions over fixed-length windows.
  \item \textit{String extraction.} Enumerates string details including: quantity, average length, printable character distribution, entropy, path incidence, URLs, registry keys, and specific strings (e.g., ``MZ'').
\end{itemize}



\end{document}